# PageRank in Malware Categorization

BooJoong Kang, Suleiman Yerima, Kieran McLaughlin, Sakir Sezer
Queen's University Belfast
Northern Ireland Science Park, Queen's Road, Queen's Island,
Belfast, Northern Ireland, United Kindom, BT3 9DT
+44 (0) 28 9097 1745
{b.kang, s.yerima, kieran.mclaughlin, s.sezer}@qub.ac.uk

## ABSTRACT
In this paper, we propose a malware categorization method that models malware behavior in terms of instructions using *PageRank*. *PageRank* computes ranks of web pages based on structural information and can also compute ranks of instructions that represent the structural information of the instructions in malware analysis methods. Our malware categorization method uses the computed ranks as features in machine learning algorithms. In the evaluation, we compare the effectiveness of different *PageRank* algorithms and also investigate bagging and boosting algorithms to improve the categorization accuracy.

## 1. INTRODUCTION
Malware detection and classification play a very important part in malware defense. Malware categorization is a malware classification technique that classifies malware into certain categories [1]. For example, multiple malware can be classified as a single malware family or a single type of malware (e.g. *Trojan*). Malware categorization can be used to discover similar malware, or groups of unknown malware, and analysts can use this additional information in further investigations on the malware. Since malware categorization can be also extended to malware detection, malware categorization can play an important role in malware defense.

Previous studies on malware categorization investigated various characteristics of malware and proposed categorization methods utilizing those characteristics. Malware is a program that consists of instructions and the instructions define the behavior of the malware. Therefore, many existing methods proposed various forms of instruction information such as instruction sequence, frequency and etc. Malware variants in the same malware family tend to reuse the original code and be written in the same development environment such as editors and compilers. Compiling the reused code with the same compiler will produce the same result, i.e. the same low-level instructions in the same structure. Some malware, which are classed as the same malware type (e.g. *Trojan*), have similar purposes and sometimes behave in the same way. Because of similar functionalities, those malware may share similar code.

Over the last few years, many research efforts have been conducted on developing automatic malware categorization systems. Various features have also been researched including instruction frequency [4-7] and sequence [8-11], control flow graph [12-14] and so on. Since a *PageRank*-based software analysis method [15] has been proposed, there is a need for investigation on *PageRank* in malware analysis.

In this paper, we propose a malware categorization method that models malware behavior in terms of instructions using *PageRank*. *PageRank* [2] is a graph-based ranking technique that computes ranks of nodes representing an importance of each node based on the structural information between nodes. A Windows-based malware can be disassembled into a set of code that consists of assembly instructions (hereafter instructions) and graphs, where a node is an instruction and an edge represents a sequence of two instructions, can be generated from the code. Ranks of instructions can be computed using *PageRank* and the ranks will be different between different malware. We investigate a number of existing *PageRank* algorithms [2-3] and compare the performance of the algorithms for malware categorization.

The remainder of this paper is organized as follows. Section 2 summarizes the related work. Section 3 describes our proposed malware categorization method with a number of existing *PageRank* algorithms. Section 4 evaluates our proposed method. Finally, Section 6 concludes the paper and outlines avenues for future work.

## 2. RELATED WORK
For many years, malware categorization has been done by human analysts but the manual analysis is time-consuming and labor-intensive [1]. Thus there has been a need for automatic malware categorization methods. One of the most essential parts in malware categorization is the feature extraction and several features have been proposed.

Since Bilar [4] discovered that the distribution of instruction frequency varies in different groups of malware, several methods have been proposed based on instruction frequency. Rad et al. [5] compute *Minkowski-form distance* of instruction frequency vectors to measure function similarities of malware variants. Ye et al. [1] applies term frequency and inverse document frequency (*TF-IDF*) to instruction frequency before clustering. Santamarta [6] extracts instruction frequency from the first 150 executed instruction as dynamic analysis in the proposed polymorphic engine classification method with neural pattern recognition. Kang et al. [7] compared the categorization accuracy of the original and repetition filtered frequencies generated from dynamic analysis. However, frequency information can be easily manipulated by obfuscation techniques.

Several efforts have also been done to use structural information as a countermeasure of obfuscation techniques. *N*-gram-based methods [8-11] have been proposed to use sequence information and the methods mostly focus on signature generation. Santos et al. [11] extract n-gram frequency of instructions for malware classification. However, n-gram-based methods will face a very high dimension feature space.

Control flow graph (*CFG*) analysis is another key commonly used in malware defense methods. A *CFG* represents a function and a program consists of multiple *CFGs*. Since maximum common sub-graph isomorphism is NP-complete, it is infeasible to compare *CFGs* directly. Gao et al. [12] use basic block comparison and Cesare el al. [13] use *edit distance* of strings generated from *CFGs*. Briones et al. [14] express a *CFG* in 3-tuple: the number of basic blocks, the number of edge and the number of sub-calls. However, *CFG* processing is time-consuming and the abstraction of *CFGs* loses good information for malware analysis.

*PageRank* [2] is a graph-based ranking technique and adopted in software analysis by Chae et al. [15] for the first time. They apply *PageRank* on *CFGs* to compute ranks of system calls for software plagiarism detection. However, there is no further research on *PageRank* in malware analysis. In this paper, we focus on instructions instead and investigate different *PageRank* algorithms for malware categorization.

## 3. PROPOSED METHOD
In this section, we explain the basic idea of *PageRank* and present two *PageRank* algorithms used in our proposed malware categorization method. We also describe how to apply *PageRank* to our malware categorization method.

### 3.1 PageRank
*PageRank* [2] is a graph-based ranking technique used to sort web pages in *Google* search engine. The basic idea of *PageRank* is that highly linked pages are more 'important' than pages with few backlinks and pages linked by highly ranked pages are more 'important' than pages linked by less 'important' pages. Based on this idea, *PageRank* gives a page a high rank if the sum of the ranks of its backlinks is high. Figure 1 shows an example of *PageRank*. In the graph of Figure 1, nodes and edges represent pages and links, respectively. Backlinks of a page $x$ are the pages that have a link to the page $x$. The page $E$ has more backlinks than the page $D$ but the page $A$, which has a link to the page $D$, is a more 'important' page than the other pages $B$ and $C$. Therefore, *PageRank* considers the page $D$ more 'important' than the page $E$. The rank of a page $x$, $PR(x)$, is given as follows:

$$PR(x) = (1-d) + d \sum_{v \in B(x)} \frac{PR(v)}{L(v)} \quad (1).$$

$B(x)$ denotes the set of backlinks of the page $x$ and $L(v)$ denotes the number of links of page $v$. The parameter $d$ is a damping factor which is proposed for the random surfer model. The parameter $d$ is the likelihood of users following the links instead of random jumps and can be set between 0 and 1. Lower values of $d$ denote more random jumps and vice versa.

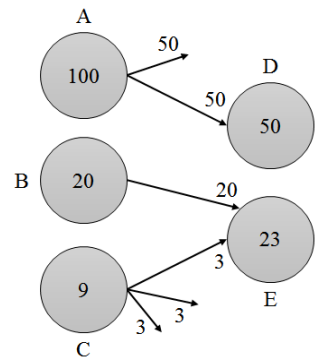

**Figure 1. PageRank**

There is a modified algorithm of *PageRank*, called *PageRank* based on visits of links (*VOL*) [3], which assumes that highly visited links are more 'important' than links with few visits. In *VOL*, the rank of a page is divided differently based on the number of visits for each link while the rank is evenly divided in the original *PageRank*. Figure 2 shows an example of *VOL* where the underlined number is the number of visits. The page $D$ gets 20 (= 2 * 100 / 10) from the page $A$ because users only visit twice on this link. *VOL* considers the page $E$ to be more 'important' than the page $D$ while *PageRank* considered the page $D$ more 'important'.

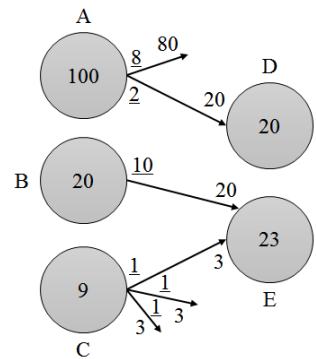

**Figure 2. Visits of Links**

In *VOL*, the rank of a page $x$, $VOL(x)$, is given as follows:

$$VOL(x) = (1-d) + d \sum_{v \in B(x)} \frac{V(x)VOL(v)}{TV(v)} \quad (2).$$

$V(x)$ denotes the number of visits of the link from the page $v$ to the page $x$ and $TV(v)$ denotes the total number of visits of all links from the page $v$.

### 3.2 PageRank in Malware Categorization
Just as web pages can be structured by links, instructions similarly construct a program by control flows. Ranks of instructions can be computed by *PageRank* and the computed ranks will be different if control flows are different because the rank highly depends on the structural information. This is the basic idea of our malware categorization method. A graph, where nodes represent instructions and edges represent control flows, can be generated from a malware sample and ranks of instruction can be computed by *PageRank* based on the generated graph. Malware of the same malware type or family is expected to show a similar distribution of ranks by *PageRank* and vice versa. Figure 3 shows an example of an instruction sequence and its graph for *PageRank*.

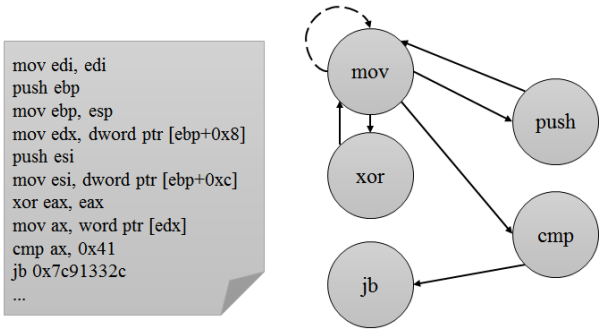

**Figure 3. PageRank of Instruction Sequence**

A node is created for each instruction and an edge is created for each *2-gram* of instructions in the instruction sequence, such as '*mov-push*', '*push-mov*' and so on. A *2-gram* of instructions is a pair of two consecutive instructions in the instruction sequence and represents a control flow of the two instructions. By creating edges in this manner, the entire control flows of the instruction sequence can be represented in a graph. When a single instruction is repeated and constructs a *2-gram* like '*mov-mov*', a self-visit edge can be created as drawn as a dashed arrow. This self-visit information is not considered in the existing *PageRank* algorithms and we will investigate the impact of the self-visit information in the evaluation. As shown in Figure 4, the number of visits can be counted for *VOL*. The two equations (1-2) can be used to compute ranks of instructions without any modification on the algorithm.

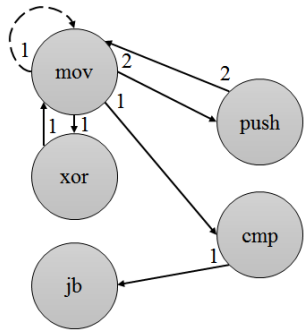

**Figure 4. VOL of Instruction Sequence**

Instructions can be extracted in a static or dynamic approach. In the static approach, a disassembler can be used to translate machine code into assembly language but packing techniques make the translation hard or impossible. In the dynamic approach, a malware sample can be executed in a controlled environment such as emulation, instrumentation or debugging. Packing might not be an issue in the dynamic approach, but anti dynamic techniques also exist such as anti-debugging, anti-virtual-machine techniques. In our proposed method, instrumentation is chosen to defeat packing techniques. As a result of instrumentation, a sequence of executed instructions can be extracted from each malware and *PageRank* algorithms can be applied to the graph which is generated from the sequence of executed instructions.

After computing ranks of instructions, we can represent the ranks in a vector as an instance of a single malware sample. By processing the above process on every malware sample, we can get rank vectors and use the vectors as input of machine learning algorithms in our proposed malware categorization method.

## 4. EVALUATION

In this section, we evaluate the *PageRank* algorithms: the original *PageRank* (*PR*) and *VOL*. We also investigate the impact of the self-visit information in the both algorithms. We use 6,721 malware samples from *VxHeaven* [16] that consist of 26 malware families and 9 malware types as shown in Table 1.

**Table 1. Malware Samples**

| Type | Family | # of Variants |
|---|---|---|
| Backdoor | Agobot | 296 |
| | Aimbot | 93 |
| | Bifrose | 148 |
| | Hupigon | 334 |
| | IRCBot | 140 |
| | PcClient | 66 |
| | Rbot | 1,141 |
| | SdBot | 863 |
| Trojan | Dialer | 194 |
| | StartPage | 216 |
| Trojan-Downloader | Banload | 321 |
| | Dadobra | 211 |
| | Dyfuca | 77 |
| | INService | 113 |
| | IstBar | 235 |
| | Swizzor | 95 |
| Trojan-PSW | LdPinch | 116 |
| | Lmir | 458 |
| | Nilage | 108 |
| | QQPass | 136 |
| Trojan-Spy | Bancos | 109 |
| | Banker | 754 |
| Email-Worm | Bagle | 128 |
| IM-Worm | Kelvir | 105 |
| Net-Worm | Mytob | 107 |
| P2P-Worm | SpyBot | 157 |
| Total | | 6,721 |

For each malware, we traced the execution of malware using *Inter Pin* [17] and extracted executed instructions, up to five million instructions. As shown in Figure 3, an instruction consists of a single opcode and multiple operands and some instructions include prefixes, such as 'lock' and 'rep', or no operand. We only extracted mnemonics, such as 'mov' and 'push', from each instruction and generated a sequence of mnemonics which are executed by the investigated malware. We generated a graph for each malware using its executed mnemonic sequence and we applied *PageRank* algorithms on each generated graph and compared the performance of malware type and family classification between the *PageRank* algorithms using *random forest* (*RF*) [18] which showed good performance in several previous work [7]. We also investigated three meta-classifiers to improve the classification accuracy: *bagging* [19], *AdaBoostM1*

[20] and *MultiBoostAB* [21]. We utilized *WEKA* **Error! Reference source not found.** to use the machine learning algorithms in the evaluation and used *f-measure*, which is the harmonic mean of *recall* and *precision* **Error! Reference source not found.**, to compare the classification accuracy. *10-fold cross-validation* is also used.

First, we evaluate the impact of the parameter *d* using *RF*. Figure 5 shows the accuracy of malware categorization: (a) malware type classification and (b) malware family classification. *SPR* and *SVOL* utilize the self-visit information while *PR* and *VOL* ignore it. We set the number of trees to 10 in *RF*. As shown in Figure 5, the classification accuracy tends to be increased as the parameter *d* is increased. As described in section 3.1, the parameter *d* is the likelihood of users' following the links instead of random jumps which are "interrupts" in the program execution. Therefore, this tendency tells us that "interrupts" were rare in the malware executions. In most cases of our evaluation, it showed the best classification accuracy when we use 1 as the parameter *d*.

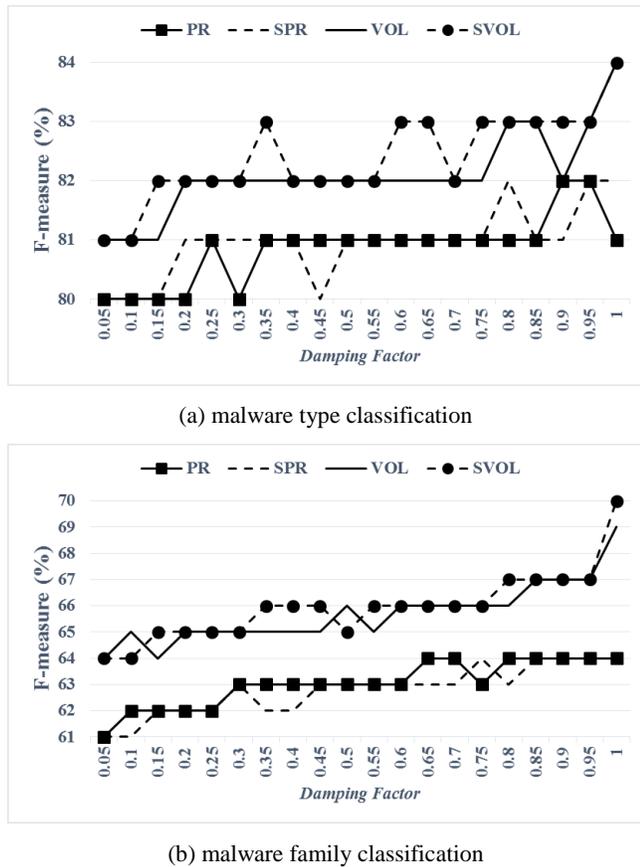

(a) malware type classification

(b) malware family classification

**Figure 5. Malware categorization with Random Forest**

We also investigated bagging and boosting algorithms to improve the classification accuracy of *RF*. The bagging and boosting algorithms will repeat the tree construction of *RF* to construct better trees. *Bagging* iterates classifier learning with different training sets, called bags, to reduce variance. In the evaluation, the size of bag and the number of iterations are set to 100% and 10, respectively. *Boosting* also reduces bias primarily and variance by iterating classifier learning and focusing on misclassified instances in every iteration. *AdaBoostM1* boosts a nominal class classifier and *MultiBoostAB* is an extension of *AdaBoost*, combining wagging's superior variance reduction. In the both boosting algorithms, the number of iterations is set to 10. We used default settings of *WEKA*.

*MultiBoostAB* shows the best classification accuracy in most cases. *SVOL* is the best *PageRank* algorithm which shows the best classification accuracy with *MultiBoostAB*: 85% in malware type classification and 72% in malware family classification.

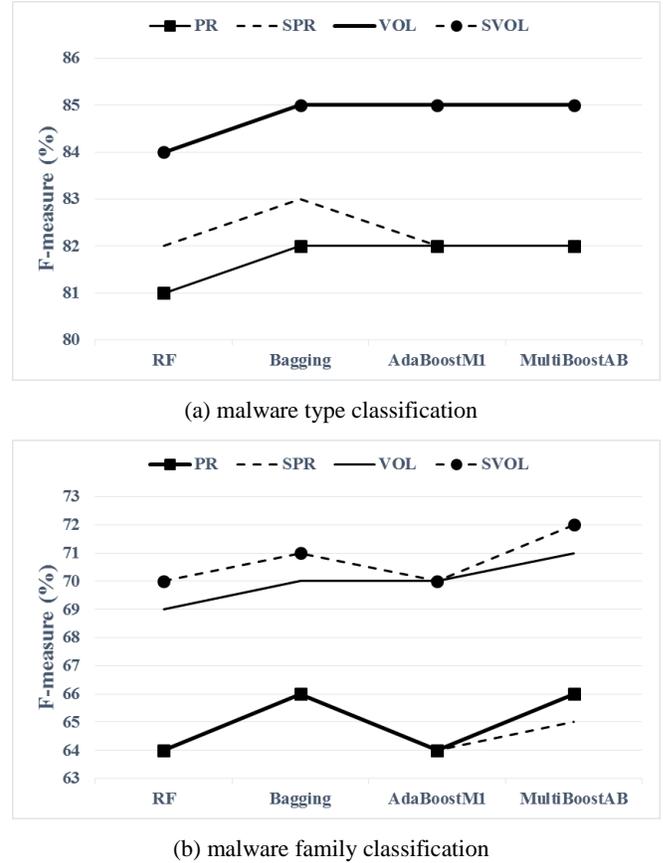

(a) malware type classification

(b) malware family classification

**Figure 6. Bagging and Boosting**

Table 2 shows the elapsed time for training and testing in *SVOL*. Bagging and boosting algorithms consumed more time compared to *RF*. It was a few seconds in this evaluation but it might not be fast enough in real situations. We need to decide which one is more important between the accuracy and the speed when we choose a good classifier.

**Table 2. Training and Testing Time (sec.)**

| Classifiers | Malware Type Classification | Malware Family Classification |
|---|---|---|
| *RF* | 0.69 / 0.0 | 0.98 / 0.0 |
| *Bagging* | 5.60 / 0.03 | 7.97 / 0.04 |
| *AdaBoostM1* | 6.56 / 0.03 | 8.18 / 0.03 |
| *MultiBoostAB* | 5.88 / 0.03 | 8.92 / 0.04 |

## 5. CONCLUSION

In this paper, we proposed *PageRank*-based malware categorization method that classifies malware types or families. First, we described how to utilize *PageRank* algorithms in malware categorization. A graph can be constructed from a sequence of executed instructions and ranks of the executed instructions can be computed by applying *PageRank* algorithms on the constructed graph. We can use the computed ranks as features in machine learning algorithms for malware categorization. In the evaluation, *MultiBoostAB* with *SVOL* showed the best classification accuracy but *RF* with *SVOL* also showed good classification accuracy and speed. This result shows that the use of the self-visit information can improve the classification accuracy of malware categorization. As future work, we will extend our proposed method to malware classification and investigate different machine learning algorithms.